\documentclass[10pt, draftclsnofoot, twocolumn]{IEEEtran}
\pdfoutput=1
\usepackage{amssymb}
\usepackage{algorithm}
\usepackage{algpseudocode}
\usepackage{algorithmicx}
\usepackage{amsfonts}
\usepackage{amsmath}
\usepackage{amssymb}
\usepackage{amsthm}
\usepackage{array}
\usepackage{graphicx}
\usepackage{bbding}
\usepackage{bigints}
\usepackage{booktabs}
\usepackage{cite}
\usepackage{geometry}
\usepackage{color}
\usepackage{diagbox}
\usepackage{epsfig,latexsym}
\usepackage{epstopdf}
\usepackage{graphicx}
\usepackage{fancyhdr}
\usepackage{float}
\usepackage{indentfirst}
\usepackage{lastpage}
\usepackage{makecell}
\usepackage{mathtools}
\usepackage{multirow}
\usepackage{pifont}
\usepackage{psfrag}
\usepackage{setspace}
\usepackage{stfloats}
\usepackage[colorlinks,linkcolor=blue]{hyperref}
\hypersetup{hidelinks,
	colorlinks=true,
	allcolors=black,
	pdfstartview=Fit,
	breaklinks=true}
\usepackage{cleveref}
\renewcommand{\baselinestretch}{1.0}
\usepackage{empheq}
\usepackage{enumerate}
\usepackage{times}
\usepackage{subfigure}
\newcolumntype{C}{>{\centering\arraybackslash$}p{\linewidth}<{$}}

\newtheorem{theorem}{Theorem}

\newtheorem{lemma}{Lemma}

\newtheorem{corollary}{Corollary}

\newcommand{\RNum}[1]{\uppercase\expandafter{\romannumeral #1\relax}}

\begin{document}

\title{Multi-Functional Chirp Signalling for Next-Generation Multi-Carrier Wireless Networks: Communications, Sensing and ISAC Perspectives}
\author{{Zeping Sui}, {\em Member,~IEEE}, {Qu Luo}, {\em Member,~IEEE}, {Zilong Liu}, {\em Senior Member,~IEEE}, {Murat Temiz}, {\em Member,~IEEE}, {Leila Musavian}, {\em Member,~IEEE}, {Christos Masouros}, {\em Fellow,~IEEE}, {Yong Liang Guan}, {\em Senior Member,~IEEE}, {Pei Xiao}, {\em Senior Member,~IEEE}, and {Lajos Hanzo}, {\em Life Fellow,~IEEE}
\thanks{Zeping Sui, Zilong Liu, and Leila Musavian are with the School of Computer Science and Electronics Engineering, University of Essex, Colchester CO4 3SQ, U.K. (e-mail: zepingsui@outlook.com, \{zilong.liu,leila.musavian\}@essex.ac.uk).}
\thanks{Qu Luo and Pei Xiao are with the 5G \& 6G Innovation Centre, University of Surrey, U.K. (email: \{q.u.luo, p.xiao\}@surrey.ac.uk).}
\thanks{Murat Temiz and Christos Masouros are with the Department of Electronic and Electrical Engineering, University College London, London, UK (emails: \{m.temiz, c.masouros\}@ucl.ac.uk)}
\thanks{Yong Liang Guan is with the School of Electrical and Electronic Engineering, Nanyang Technological University, 639798, Singapore (e-mail: eylguan@ntu.edu.sg).}
\thanks{Lajos Hanzo is with the Department of Electronics and Computer Science, University of Southampton, Southampton SO17 1BJ, U.K. (e-mail: lh@ecs.soton.ac.uk).}
}
\maketitle

\begin{abstract}
To meet the increasingly demanding quality-of-service requirements of the next-generation multi-carrier mobile networks, it is essential to design multi-functional signalling schemes facilitating efficient, flexible, and reliable communication and sensing in complex wireless environments. As a compelling candidate, we advocate chirp signalling, beneficially amalgamating sequences (e.g., Zadoff–Chu sequences) with waveforms (e.g., chirp spread spectrum and frequency-modulated continuous wave (FMCW) radar), given their resilience against doubly selective channels. In addition to chirp sequences, a wide range of chirp waveforms is considered, ranging from FMCW to affine frequency-division multiplexing (AFDM), to create a promising chirp multicarrier waveform. This study also highlights the advantages of such waveforms in supporting reliable high-mobility communications, as well as integrated sensing and communications (ISAC). Finally, we outline several emerging research directions for {chirp signalling} designs.
\end{abstract}
\IEEEpeerreviewmaketitle

\section{Introduction}\label{Section 1}
\begin{figure*}[t]
\centering
\includegraphics[width=\linewidth]{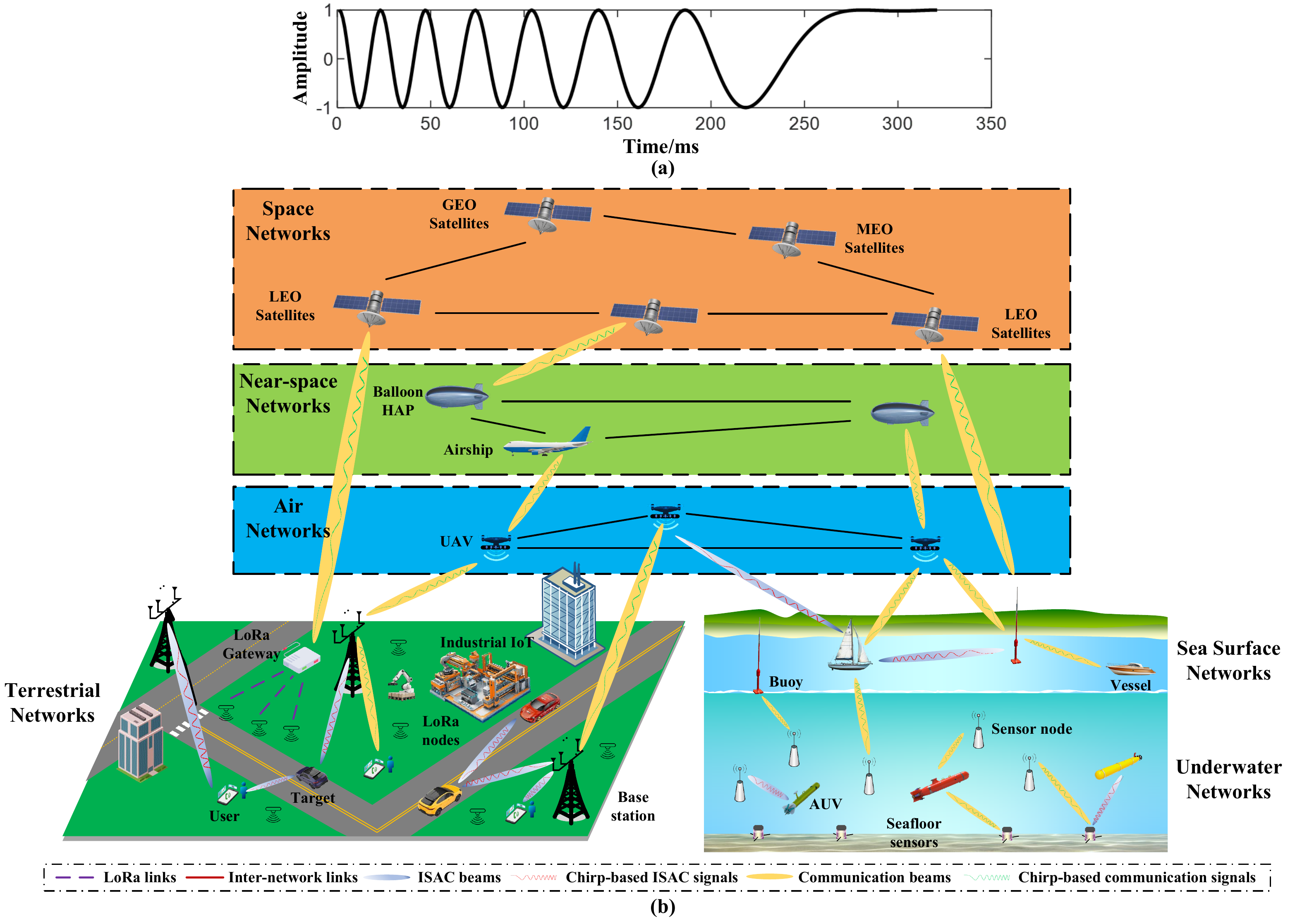}
\caption{Illustration of (a) chirp signal and (b) {chirp signals} in SAGSINs, involving non-terrestrial, terrestrial, sea surface, and underwater networks, with the integration of LoRaWAN and LEO satellites.}
\label{Figure1}
\vspace{-2em}
\end{figure*}

Chirp signalling refers to sequences and waveforms characterized by either linearly or nonlinearly varying frequency over time, as shown in Fig. \ref{Figure1}(a). While chirp signalling has recently attracted attention within new communication scenarios and sensing applications, it represents a classic concept. Since the 1960s, chirp signalling has been widely applied to communication, radar, and sonar systems due to its excellent time-frequency resolution, Doppler tolerance, and robustness to multipath \cite{klauder1960theory}. For example, chirp sequences, such as Zadoff-Chu (ZC) sequences\footnote{Discovered in the 1960s by Chu and later by Zadoff, ZC sequences are constant-amplitude and zero-autocorrelation (CAZAC) sequences designed initially for sensing and communication systems requiring high precision and low interference.}, offer excellent correlation properties, Doppler tolerance, and multipath resilience. Furthermore, frequency-modulated continuous wave (FMCW) and linear frequency-modulated (LFM) waveforms exhibit attractive Doppler-invariant properties under narrowband and short-duration conditions \cite{820736}. Furthermore, as pointed out in \cite{Lewis1982}, there is a close connection between Zadoff-Chu sequences and LFM signals. Specifically, P3 and P4 codes can be derived from a linear frequency modulation waveform, which are essentially Zadoff-Chu sequences with a root index of 1. It was found that P3 and P4 codes are more Doppler-tolerant than other phase codes. To date, these chirp signals have been widely used as preamble waveforms for Doppler estimation, synchronization, and target detection over doubly selective channels.

Subsequently, {Chirp Spread Spectrum (CSS)} solutions emerged as robust modulation techniques designed for resisting interference and multipath fading, initially explored for radar and military communications. They attracted broader attention when Spread Spectrum Communications Inc. exploited the CSS technology, and were later adopted by the Long Range (LoRa) technology and the IEEE 802.15.4a concept, designed for low-power, long-range Internet of Things (IoT) communications. Based on CSS modulation, the LoRa wide area networking (LoRaWAN) philosophy emerged as a transformative paradigm for large-scale, energy-efficient connectivity in IoT applications  \cite{Ullahlora}. 

As countries race to define next-generation (NG) mobile networks, it is imperative to design advanced sequences and waveforms to meet the increasingly stringent quality-of-service requirements, terabit data rates, ultra-reliability, ultra-low end-to-end latency, massive connection density, high spectral efficiency (SE) and energy efficiency (EE), as well as significantly enhanced environmental and situational awareness. Furthermore, multi-carrier NG networks are deemed to support reliable, seamless, and ubiquitous wireless services in highly dynamic scenarios, typically involving high-speed trains and vehicles, drones and aircraft, high-altitude platforms, and low-Earth-orbit (LEO) satellites \cite{Wang20236G}. In light of this, we portray how chirp signalling can play an essential role in shaping the multi-carrier NG system design. To illustrate our vision, we consider the Space-Air-Ground-Sea Integrated Network (SAGSIN) of Fig. \ref{Figure1}(b), which depicts a network of networks involving terrestrial, non-terrestrial, sea surface, and underwater systems, where {chirp signalling} proves to be beneficial.

Specifically, chirp signalling is promising for at least the following important research directions:  
\begin{enumerate}
\item \textit{High-mobility communications}. Multi-carrier NG systems should support reliable communications at velocities of 1000 km/h or higher. In this scenario, legacy orthogonal frequency division multiplexing (OFDM) suffers from the notorious Doppler effect, resulting in frequent loss of synchronization, destroyed subcarrier orthogonality, increased inter-carrier interference (ICI), and degraded bit error rate (BER) performance. As a remedy, orthogonal chirp-division multiplexing (OCDM)\footnote{Indeed, OCDM may be regarded as a special case of AFDM by appropriately setting the parameters of the chirp subcarriers.}, orthogonal time frequency space (OTFS)\footnote{As another application of chirp sequences in multi-carrier waveforms, Zak-OTFS can also be realized by invoking generalized DAFT, yielding spread carrier waveforms exhibiting CAZAC properties \cite{11083562}.} \cite{11083562,mattu2026low}, and affine frequency-division multiplexing (AFDM)\footnote{{Let $M$ and $N$ denote the numbers of delay and Doppler bins in Zak-OTFS, respectively. When Zak-OTFS is realized using the generalized DAFT and the AFDM parameters satisfy $c_1MN$, $c_2MN$$\in\mathbb{Z}$, Zak-OTFS and AFDM may be regarded as unitarily equivalent waveforms \cite{mattu2026low}. Nevertheless, such waveform-level equivalence does not necessarily imply complete system-level equivalence across all practical dimensions, such as sensing accuracy, computational complexity, peak-to-average power ratio (PAPR), out-of-band power emission, receiver architecture, channel estimation overhead, and robustness to hardware impairments. A comprehensive multi-component, Pareto-based comparison of these aspects is beyond the scope of this article.}} \cite{10087310} have been proposed, whereby information bits are modulated onto subcarriers in different domains, and the rapidly time-varying channels are transformed into quasi-static channels associated with sparse structures, yielding better BER performance compared to OFDM. Additionally, AFDM has demonstrated convenient backward compatibility with OFDM, requiring modest modifications to existing OFDM architectures.
\item \textit{Integrated sensing and communication (ISAC)}: Recently, significant research attention has been dedicated to ISAC under multi-functional wireless systems\footnote{In some literature, localization is also regarded under the ``sensing'' umbrella. Hence, we consider ISAC in this work for ease of presentation.}. By developing a beneficial waveform for ISAC, one can share both the spectrum and hardware for communications and sensing functionalities, thereby reducing cost. Given the inherent advantages of chirp signalling in sensing, it is important to design chirp waveforms for ISAC systems \cite{TemizRadarIsac2023}. 
\end{enumerate} 

Against the aforementioned background, we aim to stimulate further research on air interface design. We critically appraise state-of-the-art chirp signalling in sensing, communications, and ISAC. Key challenges are identified, compelling solutions are proposed, and promising open directions are discussed. While we highlight the capabilities of AFDM in supporting high-mobility communications and ISAC, it should be noted that this paper covers a wider range of chirp signalling from sequences to waveforms.

This paper is organized as follows. Chirp-aided communication techniques relying on chirp sequences, chirp preambles, LoRa modulation, and multi-carrier chirp waveforms are discussed in Section \ref{Section 2}. Then, in Section \ref{Section 3} we illustrate chirp waveforms found in sensing and ISAC. Furthermore, in Section \ref{Section 4} we identify potential challenges and promising open research directions. Finally, Section \ref{Section 5} concludes the paper. 

\section{Chirp-aided Communications}\label{Section 2}
\subsection{Chirp Sequences for Wireless Communications}\label{Section 2-1}
Chirp sequences are constant-amplitude vectors that may be regarded as discrete versions of LFM waveforms. Besides P3 and P4 codes, the most widely known chirp sequences are ZC sequences \cite{Lewis1982}, which play a pivotal role in 4G Long-Term Evolution (LTE) and 5G New Radio (5G NR) systems as preambles, synchronization, and reference signals.

Briefly, each ZC sequence exhibits zero periodic autocorrelation sidelobes and strong Doppler resilience. The discrete Fourier transform of a ZC sequence is also a finite chirp. When appropriately configured, a set of ZC sequences having different chirp rates exhibits minimum periodic cross-correlation magnitudes, meeting the celebrated Sarwate bound. Additionally, ZC sequences having a zero correlation zone can also be obtained by modulating a common perfect ``carrier" sequence with a set of orthogonal modulating sequences \cite{Popovic2010}. 

In both 4G LTE and 5G NR, ZC sequences form the basis of downlink synchronization signals, as detailed in 3GPP TS 36.211 and 3GPP TS 38.211. The sharp autocorrelation peak inherent in ZC sequences enables precise timing acquisition, which is crucial for establishing and maintaining reliable communication links. The Demodulation Reference Signals in LTE and NR also employ ZC-based sequences, leveraging their constant-amplitude property for accurate channel impulse response estimation. Furthermore, in 5G NR, ZC sequences are employed as preamble sequences in the Physical Random Access Channel by exploiting their low cross-correlation. 

Nevertheless, a comprehensive understanding of the ambiguity properties of ZC sequences remains incomplete\footnote{{It is noted that there are two typical types of ambiguity functions: 1) the periodic ambiguity function; and 2) the aperiodic ambiguity function. While the behavior of the periodic ambiguity function of CAZAC sequences is well understood, it is relatively challenging to analyze the aperiodic case.}}. Excellent ambiguity properties are required for supporting integrated sensing, localization, and communications in dynamic environments. However, due to their limited set size constrained by their unique features, ZC sequences alone may not be able to support efficient random access in massive machine-type communication systems. Therefore, it would be interesting to explore specifically structured supersets as an alternative to ZC sequences.

\subsection{Chirp Preamble Waveforms for Underwater Acoustic Communications}\label{Section 2-3}
\begin{figure}[t]
\centering
\subfigure[]{\label{Figure2-1}\includegraphics[width=0.9\linewidth]{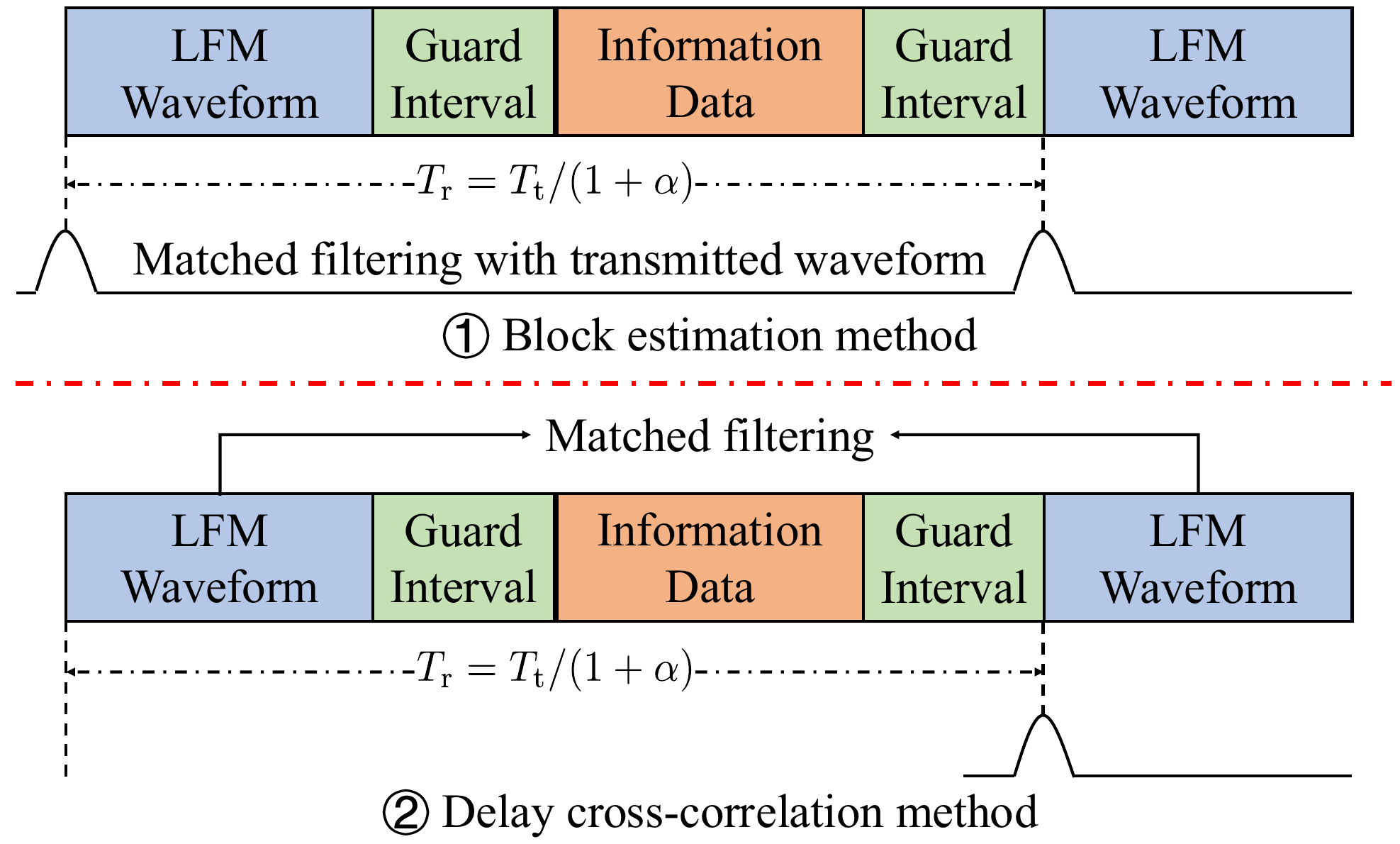}}
\vskip -0.5em
\subfigure[]{\label{Figure2-2}\includegraphics[width=0.9\linewidth]{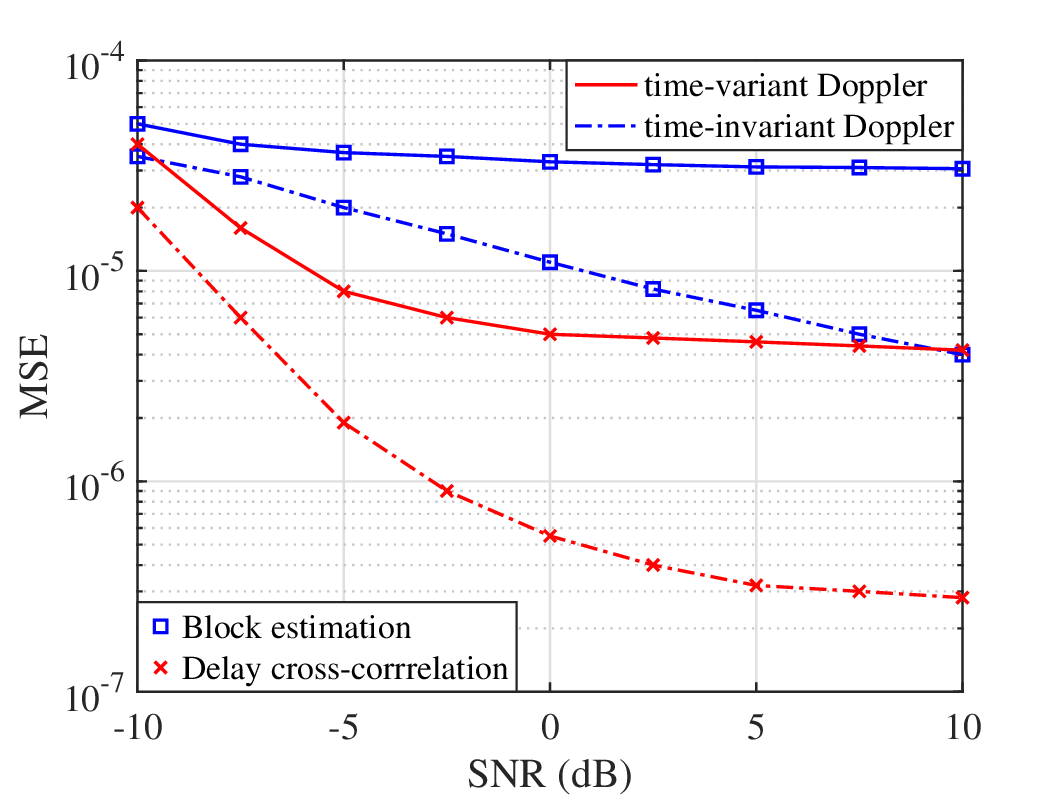}}
\caption{(a) Frame structure of chirp preamble waveform-based Doppler estimation schemes; (b) MSE performance of chirp preamble waveforms for Doppler estimation operating with both time-variant and time-invariant Doppler shifts.}
\label{Figure2}
\vspace{1em}
\end{figure}

Underwater acoustic communications (UAC) can outperform its radio and optical counterparts, since acoustic waves may propagate over longer underwater distances. Hence, they can support mission-critical applications, such as oceanic monitoring, underwater surveillance, and autonomous underwater swarm coordination. However, UAC systems face major technical challenges, where the Doppler effects are particularly severe. Due to the low speed of sound in water and the relative motion of transmitters, receivers, and the medium itself, even small movements can cause significant Doppler shifts and spreads. The dispersion and inter-symbol interference severely degrade communication reliability \cite{820736}.

Here we introduce two chirp waveform-based methods for Doppler estimation in UAC, namely block estimation and delay cross-correlation, whose frame structures are illustrated in Fig. \ref{Figure2-1}. Here, $T_t$ and $T_r$ denote the time interval between two chirp waveforms at the transmitter and receiver, respectively, while $\alpha$ is the Doppler scaling factor. Notably, the chirp waveforms used for Doppler estimation are continuous, and each frame in Fig. \ref{Figure2-1} is formed by concatenating different continuous time-domain signal components. By contrast, the vector-based ZC sequences used in LTE and 5G NR systems are discrete.

Both the block estimation and delay cross-correlation methods exploit the fact that Doppler shifts stretch or compress the length of received data frames. In the block estimation method, Doppler-resistant LFM preambles are inserted, and the Doppler shift is determined by cross-correlating the transmitted preambles with their received counterparts and comparing them to the signal frame lengths. The delay cross-correlation method, on the other hand, computes the cross-correlation between a chirp-based prefix and a padding chirp waveform. The block estimation and delay cross-correlation methods offer low computational complexity. However, the block estimation method may produce multiple cross-correlation peaks associated with random amplitudes in time-varying multipath channels, reducing its reliability. Moreover, it assumes a constant Doppler shift within a frame, leading to poor estimation performance when the Doppler factor varies significantly over longer durations.

In Fig. \ref{Figure2-2}, we compare the Doppler estimation performance outlined above under both time-invariant and time-variant Doppler shift scenarios, using parameters set as in \cite{820736}. One can see that the mean squared error (MSE) performance of all the methods degrades under time-variant scenarios. Moreover, since the channel variation becomes more pronounced over two chirp waveforms, a significant performance erosion is observed for the block estimation method compared to that of the delay cross-correlation method.

\subsection{LoRa Modulation with Chirp Spread Spectrum}\label{Section 2-2}
\begin{figure*}[htbp]
\centering
\includegraphics[width=\linewidth]{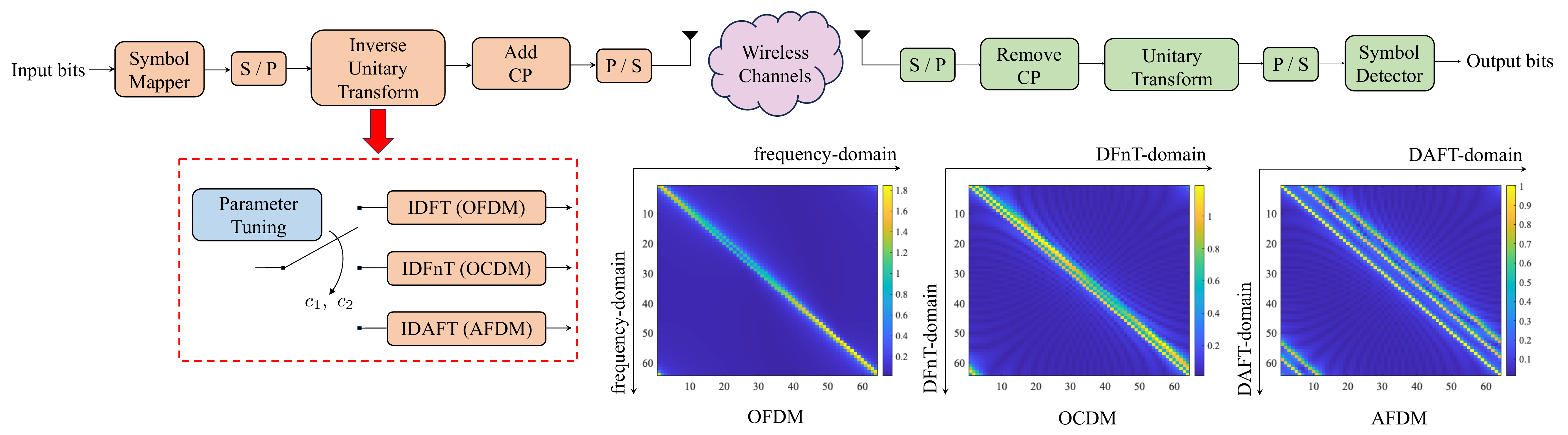}
\vspace{-1em}
\caption{Illustration of different modulation schemes and the three-path effective frequency-, discrete Fresnel transform (DFnT)-, and DAFT-domain effective channel matrices of OFDM, OCDM, and AFDM, i.e., the horizontal and vertical axes of these matrices are the indices of channel matrix elements in frequency-, DFnT-, and DAFT-domain, respectively. The normalized delay and Doppler profiles are exploited as $\mathcal{L}=\{0,1,2\}$ and $\mathcal{K}=\{0.2,1.3,1.4\}$ with $64$ subcarriers.}
\label{Figure4}
\vspace{-2em}
\end{figure*}

Based on the CSS modulation technique, LoRa provides low-power and long-range wireless communications, especially for IoT networks. LoRa operates in the sub-GHz unlicensed industrial, scientific, and medical band, using specific frequency ranges depending on the region, e.g., $863$–$870$ MHz in Europe and  $902$–$928$ MHz in the USA. It supports multiple channel bandwidths, including $125$ kHz, $250$ kHz, and $500$ kHz. A key configurable parameter in LoRa modulation is the spreading factor (SF), which determines the number of chirps used per symbol and directly influences the data rate, robustness, and range. The spreading factor is defined as $\text{SF} = \log_2 M$, where $M$ denotes the modulation order. In practice, LoRa systems typically utilize SF values ranging from $6$ to $12$. Higher SFs (e.g., $10$–$12$) result in longer chirps, which exhibit increased processing gain and sensitivity, supporting longer-range but lower-rate communications. Conversely, lower SFs (e.g., $7$–$9$) allow higher data rates at the expense of reduced communication range. Since LoRa defines the PHY layer of the communication stack, it may have diverse network protocols, and LoRaWAN is the most popular one. Hence, LoRaWAN is widely regarded as the MAC protocol for LoRa, which mainly defines the network architecture and its bi-directional communication protocol.

Owing to its long-range transmission capability and ultra-low power consumption, LoRa has been widely adopted in terrestrial applications such as wireless sensor networks and smart cities. Recently, the integration of LEO satellites with LoRa-based IoT systems has attracted increasing interest in both academia and industry, aiming to extend coverage to underserved remote regions. LoRa communication in LEO systems can be broadly categorized based on gateway deployment. In the Gateway-on-Satellite model, the LoRa gateway functionality is embedded onboard the satellite, facilitating direct device-to-satellite connectivity with global reach. This approach provides wider coverage but requires sophisticated onboard processing to manage Doppler shifts and protocol compatibility. Lacuna Space has demonstrated the feasibility of this model by deploying five LEO satellites equipped with onboard LoRa gateways capable of successfully receiving LoRa messages from ground terminals \cite{Ullahlora}. By contrast, the Gateway-on-Ground model treats satellites as transparent relays, forwarding signals to terrestrial gateways. While this method reduces satellite complexity and cost, it is constrained by the availability of ground infrastructure and satellite visibility windows.

\subsection{Multicarrier Chirp Waveforms for Communications}\label{Section 2-4}
\begin{figure}[t]
\centering
\subfigure[]{\label{Figure5-1}\includegraphics[width=0.9\linewidth]{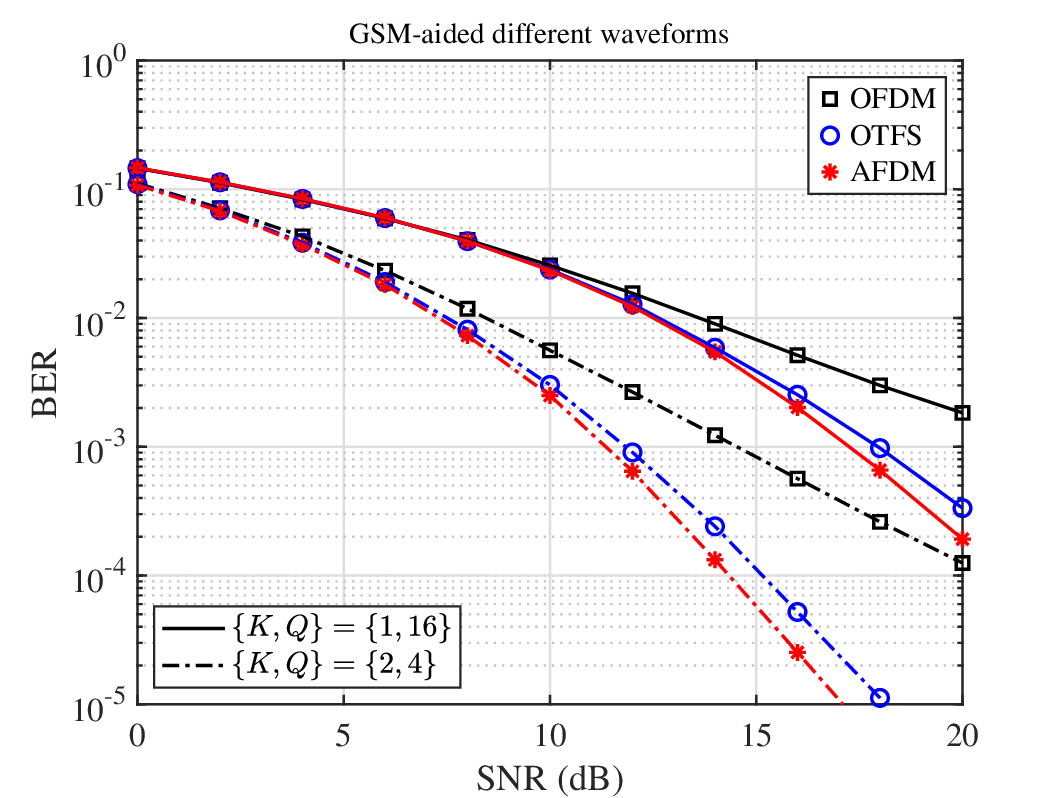}}
\vskip -0.5em
\subfigure[]{\label{Figure5-2}\includegraphics[width=0.9\linewidth]{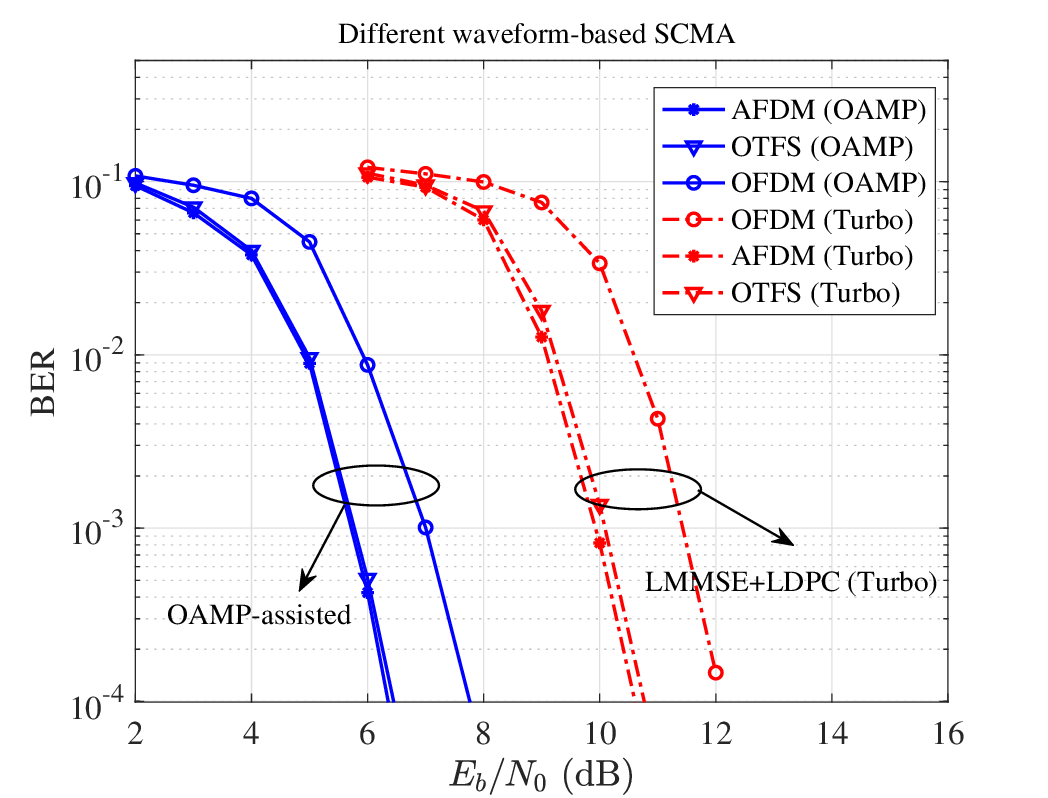}}
\caption{BER performance comparison of \subref{Figure5-1} GSM-OFDM, GSM-OTFS, and GSM-AFDM operating in four-path channels with the carrier frequency $f_c=4$ GHz, carrier spacing $\Delta f=2$ kHz and maximum velocity $v=400$ km/h, as well as \subref{Figure5-2} downlink coded OFDM-SCMA, OTFS-SCMA, and OFDM-SCMA through the EVA channel using carrier frequency $f_c=4$ GHz, carrier spacing $\Delta f=15$ kHz and maximum velocity $v=300$ km/h, where the remaining system parameters can be found in \cite{sui2025generalized} and \cite{10566604}, respectively.}
\label{Figure5}
\end{figure}

Multicarrier chirp waveforms, such as OCDM and AFDM, that can support reliable high-mobility communications were introduced in Section \ref{Section 1}. In contrast to the conventional OFDM waveform, which exhibits linear phase evolution across a symbol, in OCDM/AFDM, every data symbol is modulated over a chirp subcarrier. From a BER perspective, AFDM generally outperforms OCDM as it may achieve higher diversity by appropriately tuning its chirp rate according to the channel's Doppler profiles. In Fig. \ref{Figure4}, we plot the structures of effective channel matrices of OFDM, OCDM, and AFDM in the frequency-, DFnT-, and DAFT-domain, respectively. As shown in Fig. \ref{Figure4}, AFDM is designed based on the discrete affine Fourier transform (DAFT) and can be regarded as a generalization of OFDM and OCDM. The AFDM parameter $c_1$ determines the chirp rate, while $c_2$ is associated with initial phases. Consequently, $c_1$ and $c_2$ can be dynamically adjusted based on the specific communication channel to achieve the best possible performance even under doubly-dispersive channels. Furthermore, AFDM can be efficiently implemented by modest modifications of the existing OFDM architecture, thus demonstrating convenient backward compatibility.

\begin{figure*}[htbp]
    \centering
{\includegraphics[width=0.8\linewidth]{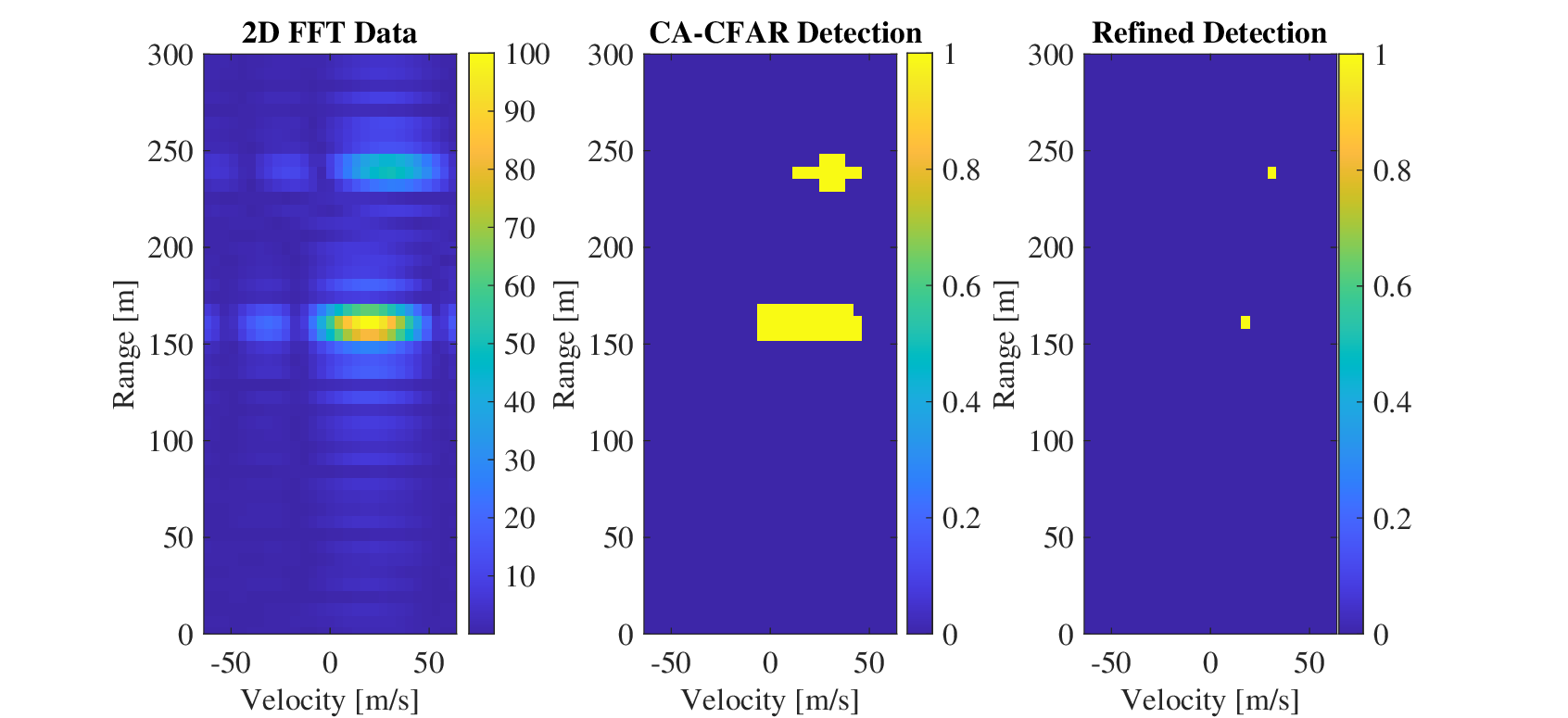}}
    \caption{Multi-target estimation parameter estimation using AFDM signals by employing CA-CFAR detection and refined detection techniques. AFDM signals consisting of $L=128$ subcarriers modulated with $64$-QAM over $B=20$ MHz bandwidth are considered, and $64$ AFDM symbols are processed to obtain the range-velocity images of two targets.
    }
    \label{fig:multi_target}
    \vspace{-1em}
\end{figure*}

The three-path effective channel responses of OFDM, OCDM, and AFDM are shown in Fig. \ref{Figure4}. Observe that many of the non-zero elements in the frequency-domain channel matrix of OFDM are significantly overlapped, while they are only partially overlapped in OCDM. This is because OFDM cannot attain multi-path diversity in doubly selective channels by using only IDFT. Moreover, the chirp parameters of OCDM are fixed, hence it cannot fully avoid overlapping among different paths. This explains why the BER performance of OFDM suffers significantly due to ICI and why AFDM achieves higher diversity than OCDM. In contrast, by carefully adjusting the chirp parameters according to the specific Doppler shifts encountered, we can see that propagation paths with distinct delay-Doppler profiles can be fully separated in the DAFT-domain, thereby resulting in a higher diversity order for AFDM.

To further explore the potential of AFDM, we have studied generalized spatial modulation (GSM)-aided AFDM \cite{sui2025generalized}, which conveys additional information by mapping extra bits onto the indices of active transmit antennas. Hence, a lower throughput, but more robust classic constellation can be used at a given total data rate. The BER performance of GSM-OFDM, GSM-OTFS, and GSM-AFDM is provided in Fig. \ref{Figure5-1}, where $K$ and $Q$ denote the number of active transmit antennas and modulation order, respectively. We can see that GSM-AFDM and GSM-OTFS consistently achieve improved BER over conventional GSM-OFDM. Moreover, for a given pair of $\{K,Q\}$, GSM-AFDM outperforms GSM-OTFS in the high SNR region, since AFDM achieves a higher diversity order, whilst the asymptotic diversity of OTFS is one. Specifically, by considering a $4\times 4$ multiple-input multiple-output (MIMO) system having $64$ resource blocks, GSM-AFDM achieves about $1$ dB SNR gain over GSM-OTFS at a BER of $10^{-5}$ and $\{K,Q\}=\{2,4\}$. To support massive connectivity under high-mobility scenarios, an AFDM-based sparse code multiple access (SCMA) system is proposed in \cite{10566604}. The BER comparison of $2/3$-rate low-density parity-check coded downlink AFDM-SCMA and its counterparts using an orthogonal approximate message passing receiver is shown in Fig. \ref{Figure5-2}. Observe from Fig. \ref{Figure5-2} that the AFDM-SCMA proposed achieves about $1.5$ dB SNR gain over OFDM-SCMA. Furthermore, AFDM-SCMA attains a similar BER performance to that of AFDM-OTFS.

\section{Chirp-aided Sensing and ISAC}\label{Section 3}

\begin{figure*}[htbp]
\centering
\includegraphics[width=0.8\linewidth]{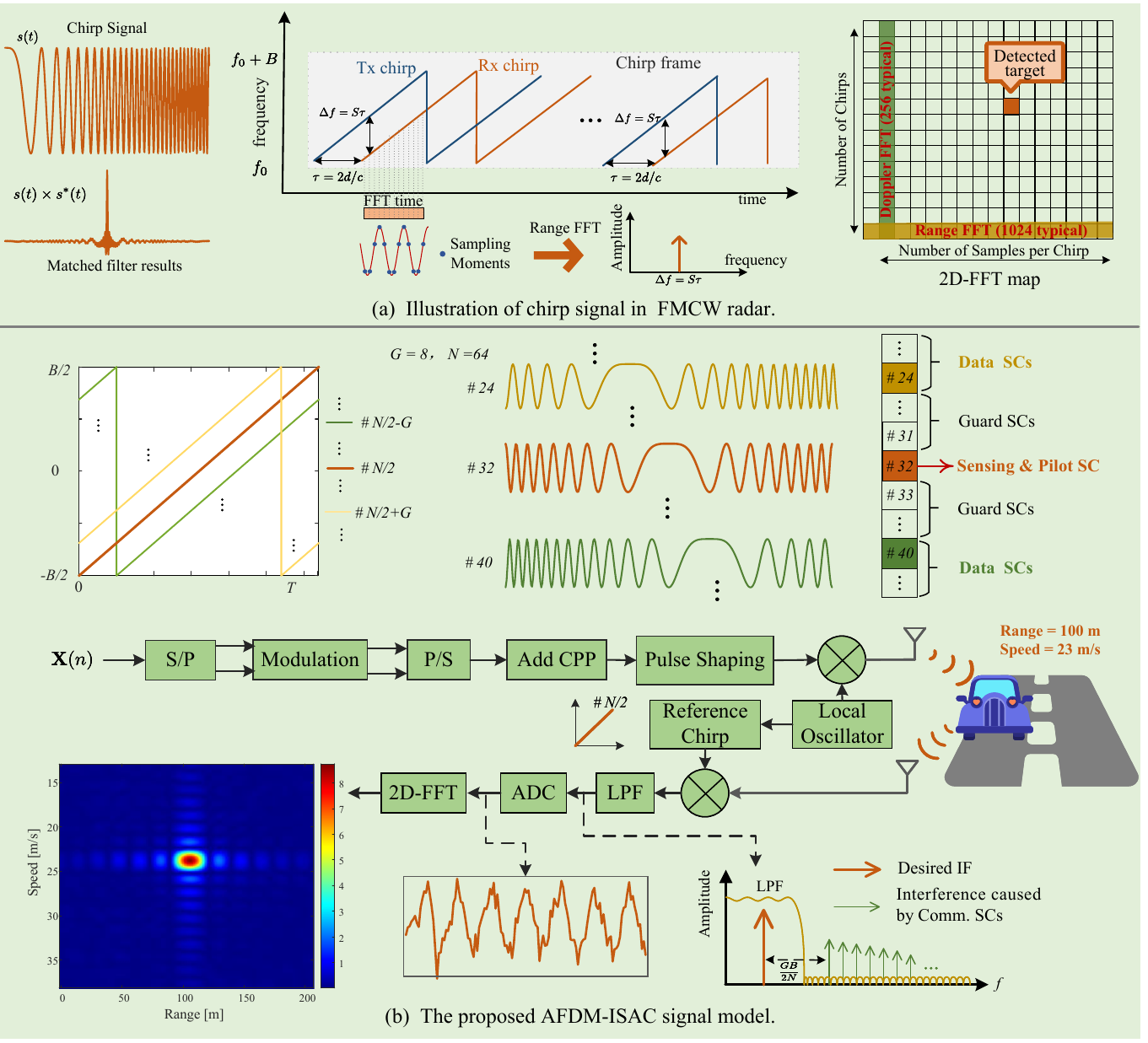}
\caption{Illustration of (a) chirp signal in FMCW radar and (b) the proposed AFDM-ISAC scheme. Here, $S$ denotes the chirp rate, $\tau$ denotes the delay, $B$ denotes the bandwidth, $N$ is the number of AFDM subcarriers, and $G$ is the length of the guard band. SC: subcarrier;  CPP: chirp-periodic prefix; P/S: parallel-to-serial conversion;  S/P: serial-to-parallel conversion; ADC: analog-to-digital converter; LPF: low-pass filter. }
\label{ISAC}
\vspace{-2em}
\end{figure*}
 
\subsection{Chirp Waveforms in Sensing}

{Chirp}-based waveforms, such as FMCW and LFM, can achieve superior target detection and precise distance estimation due to their high time-bandwidth products attained through chirp pulse compression. Fig. \ref{ISAC}(a) illustrates the signal processing flow of a matched-filter-assisted FMCW radar. Specifically, by sending a chirp waveform that sweeps over a certain bandwidth, the Doppler frequency shift can be observed as a phase change across chirps. Joint range-Doppler processing is typically conducted via a two-dimensional fast Fourier transform (2D-FFT). As seen, an FMCW radar compresses the long-duration chirp waveform into a much shorter impulse-like waveform, thereby enhancing range resolution while maintaining energy efficiency.  Nevertheless, FMCW radar may suffer from performance degradation in multipath environments, as indirect propagation paths can produce extra frequency components of the mixed signal that interfere with the true target responses. Consequently, false or ambiguous peaks may appear, thereby reducing the reliability of target detection and the accuracy of range and velocity estimation.

Note that the range resolution of radar systems employing noncoherent detection is primarily determined by the signal bandwidth. Hence, FMCW radar systems targeting short-range applications often harness wideband chirps at millimeter-wave (mmWave) carrier frequencies to achieve precise range resolution, accurately distinguishing small distances between targets. However, given the high path loss of the twin-hop mmWave signal, the maximum range becomes limited. By contrast, the Doppler resolution depends on the coherent processing interval.  Additionally, chirp waveforms are resilient to multipath fading due to their low autocorrelation sidelobes and can be efficiently synthesized using low-complexity digital signal processing techniques. These appealing properties make chirp waveforms eminently suitable for sensing systems. 

Multi-target estimation can be performed by processing the range-velocity images obtained by the radar. Fig. \ref{fig:multi_target} illustrates the target parameter estimation performance for a pair of targets using the AFDM waveform, where the constant false-alarm rate cell-averaging (CA-CFAR) algorithm is used for the detection of targets relying on the range-velocity image, which is formed by applying a fast-time range FFT followed by a slow-time Doppler FFT to the data obtained after the matched-filtering\footnote{This consecutive range FFT and Doppler FFT processing is mathematically equivalent to a two-dimensional fast Fourier transform (2D FFT).}. Then, these detection operations are refined by retaining only the strongest peak for each target, followed by estimating the targets' ranges and velocities. Note that AFDM signals are modulated with communication data such that the same waveform is used for both sensing and communications in this scenario.


\subsection{Multicarrier Chirp Waveforms for ISAC}
Given their advantages in both sensing and communications, chirp waveforms constitute an excellent waveform candidate for ISAC. Under a multicarrier architecture, AFDM enables a transmitter to illuminate targets and transmit high-rate data simultaneously using a common waveform. The chirps reflected from targets can be processed using techniques such as ``dechirping" or DAFT processing to extract target range and velocity information. A notable feature of AFDM is its ability to extract sensing information from even a single pilot chirp embedded in a transmission frame. This characteristic, inherent in the chirp structure, permits efficient estimation of multipath delays and Doppler shifts at a modest overhead \cite{bemani2024integrated}.

A powerful AFDM-ISAC scheme that fully exploits these inherent chirp advantages is shown in Fig. \ref{ISAC}, which uses pulse compression via dechirping. The key idea is to use a single chirp subcarrier for high-resolution sensing. As shown in Fig. \ref{ISAC}(b), the AFDM-ISAC waveform utilizes the $N/2$-th chirp subcarrier for sensing, along with double guard bands of length $G$ on both sides. At the receiver, dechirping is applied to that chirp subcarrier. By carefully designing a low-pass filter, the interference imposed by communication subcarriers can be effectively eliminated, both at the transmitter and in the received echoes. 
For monostatic sensing, where the ISAC transmitter and receiver are co-located, the receiver suffers from strong onboard leakage from the transmitter, which necessitates self-interference cancellation due to full-duplex operation. In the proposed design, mixing the sensing chirp with the onboard transmitter leakage yields only a direct-current component, whereas the other subcarriers produce much higher-frequency components than the target. Hence, the latter may be readily suppressed by the low-pass filter (LPF). By placing the analog-to-digital converter (ADC) after the LPF, receiver saturation induced by transmitter leakage can be effectively mitigated. As a result, the proposed scheme eliminates the need for expensive full-duplex RF isolation, thereby enabling simultaneous transmission and reception at a relatively low hardware complexity.

{Note that the above receiver operates in the continuous time domain and processes each chirp segment separately\footnote{The related comments of the anonymous reviewer are much appreciated.}. By contrast, as highlighted in \cite{nisar2026zak}, discrete chirps enable simultaneous resolution in both the delay and Doppler domains. Moreover, the authors of~\cite{mehrotra2025discrete} initiated the investigation of AFDM-based sensing associated with discrete chirps and demonstrated the potential of this approach for reducing both the radar-processing complexity and PAPR\footnote{{Under the unitary-equivalence conditions discussed earlier, mathematically, the corresponding advantages may also be shared by Zak-OTFS. From an algebraic perspective, the associated waveform bases can be interpreted as common eigenvectors of a maximal commutative subgroup of the Heisenberg-Weyl group, allowing the corresponding time-frequency shifts to be factored out of cross-correlation calculations.}}. Building on these initial results, further development of discrete-chirp AFDM sensing remains an interesting direction for future research.}

\section{Future Challenges and Opportunities}\label{Section 4}
\subsection{Chirp-like Sequences Beyond ZC}
For multi-functional multi-carrier NG networks, it is important to look for new chirp-like sequences that can go beyond the existing ZC sequences. To support massive machine-type communications, chirp-like supersets (e.g., each comprised of multiple small sequence sets) are desirable. Subject to the fundamental Welch bound and the Sarwate bound, new structural properties (e.g., zero or low inter-set correlation functions) may be explored. Such chirp-like supersets are also of great interest in improving the throughput and latency of random access. Additionally, one may look for new chirp-like preamble sequences having significantly enhanced ambiguity functions. Instead of having low-ambiguity sidelobes across the entire delay-Doppler domain, a promising direction is to design chirp-like sequences with ``zero/low ambiguity zone" properties. Various optimization tools (e.g., majorization-minimization method) and machine learning may be useful for this exploration.

\subsection{OCDM \& AFDM Meet Complicated Channels}
Since the chirp parameters in OCDM are fixed, it cannot obtain full diversity and lacks extra degrees of freedom to achieve flexible performance control across different applications. {On the other hand, assuming a sufficiently large bandwidth, existing studies on AFDM have mainly considered doubly selective channels having integer delays and fractional Doppler shifts.} However, fractional delays become inevitable in limited-bandwidth scenarios, and AFDM is expected to flexibly tune the bandwidth to specific applications. Both the systematic investigation of AFDM performance under fractional delays and the corresponding values of the chirp parameters $c_1$ and $c_2$ to achieve full diversity are open problems at the time of writing, and require further research. On the other hand, good BER performance for AFDM with full diversity, even in conventional integer-delay fractional-Doppler cases, highly depends on the particular value of chirp parameters, which should be carefully tuned based on the maximum Doppler shift. This approach assumes that the AFDM receiver perfectly knows $c_1$ and $c_2$. Nonetheless, the maximum Doppler shift is difficult to estimate precisely in practical scenarios, which limits the practical application of AFDM. Future research should focus on AFDM design with blind chirp-parameter estimation or on AFDM systems capable of chirp-parameter estimation.

\subsection{AFDM-based EHF Wireless Networks}
The ultra-wide bandwidth of mmWave and Terahertz signals has been exploited by numerous researchers to achieve both ultra-high data rates and ultra-low latency while improving the sensing resolution. Under this scenario, the AFDM-based communication and sensing systems suffer from severe path loss and hardware impairments such as phase noise and the non-ideal nature of analog-to-digital/digital-to-analog converters. Therefore, AFDM can be leveraged in massive MIMO systems, harnessing advanced beamforming schemes to achieve good beam alignment performance under high-mobility environments. Moreover, novel joint hardware impairment estimation and compensation methods may be developed to enhance communication and sensing performance. Furthermore, extremely high-frequency (EHF) channels are often sparse in the delay and angular domains, which matches well with the AFDM characteristics to efficiently represent sparse delay-Doppler profiles, enabling compressed sensing-based estimation and detection. 

\subsection{Chirp-based Multiple Access in NTNs}
Non-terrestrial networks (NTNs) are expected to play a vital role in multi-carrier NG systems, supporting a large number of devices as a remedy for the limitations of terrestrial systems. Considering the high mobility of LEO satellites, it is natural to investigate chirp waveform-based multiple access techniques for NTN paradigms. Explicitly, diverse services and device types are employed in NTNs, resulting in heterogeneous networks. Therefore, hybrid waveforms such as OFDM combined with AFDM can be considered. Finding efficient protocols for network management remains an open challenge for future research.
\subsection{Secure Transmission Meets Chirp Waveforms}
Given the extensive scale of device and data flows in the SAGSIN system shown in Fig. \ref{Figure1}, future research should prioritize secure transmission while maintaining ubiquitous connections. Physical layer security (PLS) presents a promising solution for chirp-based signal systems. Specifically, chirp preamble waveforms having different chirp rates and diverse AFDM parameters can be exploited by various users to enhance security. Furthermore, machine learning methods may be utilized in chirp-based multiple access systems. However, finding the optimal Pareto-front of all non-dominated solutions in terms of the BER, security, SE, and EE is an attractive but challenging topic for future research.
\subsection{Deep Learning-based Techniques for Chirp Waveforms}
Design of optimum ISAC systems using chirp waveforms involves non-convex optimization problems. Solving such non-convex problems usually requires iterative methods that are computationally expensive. Deep learning-based techniques can produce near-optimum solutions with low computational complexity and memory usage, enabling real-time implementation at transmitters and receivers. For instance, deep learning-based target detection and parameter estimation algorithms can be developed specifically for chirp waveforms to improve the sensing performance in the presence of clutter or complex sensing scenarios. Moreover, deep learning-based techniques can be used to optimize waveform and precoder design to meet the strict delay constraints of highly dynamic communication and sensing scenarios. 

\subsection{Hardware Design, Optimization and Prototyping}
To maximize the ISAC performance and energy efficiency of chirp waveforms, it is essential to design and optimize both the hardware and the algorithms specifically tailored for their unique characteristics, which differ fundamentally from conventional waveforms. For instance, reducing the peak-to-average power ratio of chirp waveforms and designing low-complexity transceivers for sensing and communications by exploiting their unique features might lead to their application in multi-carrier NG networks. Moreover, developing prototypes and performing field experiments is necessary to verify their performance under realistic channel conditions, including high-velocity-induced hardware impairments in the face of interference.

\subsection{Chirp Waveform Design With MIMO}
In both communication and sensing contexts, spatial processing plays a vital role in enhancing the angular resolution, spatial diversity, multiplexing, and interference suppression. When integrated with massive MIMO architectures, chirp waveforms such as AFDM and OCDM may significantly enhance system performance in high-mobility scenarios and doubly selective channels. From a sensing perspective, the joint design of chirp waveforms and antenna array patterns can significantly enhance angular resolution and facilitate more accurate multi-target tracking. For example, radar systems that introduce frequency offsets across antennas can be naturally combined with chirp waveforms to generate range-angle-dependent beampatterns, facilitating high-precision target tracking. In summary, the co-design of antenna architecture, waveform structure, beamforming strategies, and signal processing algorithms is essential for fully unleashing the potential of spatial-domain chirp waveforms for communication, sensing, and even ISAC in multi-carrier NG systems.

  
\section{Conclusions}\label{Section 5}
The promising multi-functional nature of {chirp signalling} designed for multi-carrier NG networks was discussed, involving terrestrial, non-terrestrial, sea surface, and underwater systems. Its robustness to Doppler shifts and time dispersion opens attractive opportunities in sensing, communication, and ISAC. We reviewed a suite of representative techniques, including ZC sequences, chirp preambles, LoRa, and FMCW radar, then we highlighted the potential of AFDM in high-mobility scenarios and ISAC. Further research directions covered the applications of AFDM in EHF bands, NTNs, secure communication, and spatial-domain waveform design. Moreover, the research opportunities associated with channel estimation and hardware prototyping were also touched upon. From an industry perspective, this article aspires to stimulate further groundbreaking research and discussion in the field, with more chirp-like sequences and chirp waveforms that can be adopted or standardized in modern wireless systems.
\renewcommand{\refname}{References}
\mbox{} 
\nocite{*}
\bibliographystyle{IEEEtran}
\bibliography{main.bib}
\end{document}